\documentclass[12pt,a4paper]{article}
\usepackage{amsmath}
\begin{document}
\title{\Large The Superposition Principle of Waves Not Fulfilled 
under M.W. Evans' $O(3)$ Hypothesis}
\author{\\Erhard Wielandt\\
Institute of Geophysics, Stuttgart University, Germany\\
ew@geophys.uni-stuttgart.de, e.wielandt@t-online.de}


\maketitle

\vspace{1cm}
\linespread{0.7}


\begin{abstract}
In 1992 {\sc M.W. Evans} proposed a so-called $O(3)$ 
symmetry of electromagnetic fields by adding a
constant longitudinal "ghost field" to the well-known 
transversal plane em waves. He considered this symmetry 
as a {\em new law of electromagnetics}. Later on, since 2002,  
this $O(3)$ symmetry became the center of his 
Generally Covariant Unified Field Theory which he recently 
renamed as ECE Theory. One of the best-checked laws of electrodynamics 
is the principle of linear superposition of electromagnetic 
waves, manifesting itself in interference phenomena. 
Its mathematical equivalent is the representation of electric and
magnetic fields as vectors. By considering the superposition of two  
phase-shifted waves we show that the superposition principle 
is incompatible with {\sc M.W. Evans}' $O(3)$ hypothesis.
\end{abstract}

\vspace*{1cm}

\section{{\sc M.W. Evans'} $ O(3) $ hypothesis}

In the following text quotations from {\sc M.W. Evans}' GCUFT book \cite{Evans 1} appear 
with equation labels [1.nn] at the left margin.\\

The assertion of $O(3)$ symmetry is at the center of {\sc M.W. Evans}' 
considerations since 1992. He claims
that each plane circularly-polarized electromagnetic wave is 
accompanied by a constant longitudinal field ${\bf B^{(3)}}$, 
a so-called "ghost field".
In addition to numerous papers, with far reaching implications as e.g. in 
\cite{Evans 2} and \cite{Evans 3}, {\sc M.W. Evans} is author of {\em five} books
on "The Enigmatic Photon" dealing with the claimed $O(3)$-symmetry of
electromagnetic fields. His hypotheses have met many objections. The reader
will find a historical overview written by {\sc A. Lakhtakia} in 
\cite[Sect.5]{Bruhn and Lakhtakia}.\\

{\sc M.W. Evans} considers a circularly polarized plane electromagnetic wave
propagating in z-direction, cf. \cite[Chap.1.2]{Evans 1}. 
Using the electromagnetic phase\\

[1.38] \hspace*{4.8cm} $\Phi = \omega t - \kappa z $,\\

\noindent
where $\kappa = \omega /c$, he describes the wave relative to his 
complex circular basis [1.41] derived from the cartesian basis vectors
{\bf i, j, k}. The magnetic field is given as\\
 
[1.43/1] \hspace*{3cm} ${\bf B}^{(1)} = \frac{1}{\sqrt{2}} B^{(0)}   
({\bf i} - i {\bf j}) e^{i \Phi} ,$\\

[1.43/2] \hspace*{3cm} ${\bf B}^{(2)} = \frac{1}{\sqrt{2}} B^{(0)}   
({\bf i} + i {\bf j}) e^{- i \Phi} ,$\\

[1.43/3] \hspace*{3cm} ${\bf B}^{(3)} = B^{(0)} {\bf k} ,$\\ 
 
\noindent
and satisfies the "cyclic $O(3)$ symmetry relations"\\

[1.44/1] \hspace*{3cm} 
${\bf B}^{(1)} \times {\bf B}^{(2)} = i B^{(0)} {\bf B}^{(3)*} ,$\\

[1.44/2] \hspace*{3cm}
${\bf B}^{(2)} \times {\bf B}^{(3)} = i B^{(0)} {\bf B}^{(1)*} ,$\\

[1.44/3] \hspace*{3cm} 
${\bf B}^{(3)} \times {\bf B}^{(1)} = i B^{(0)} {\bf B}^{(2)*} .$\\
 
Especially equ.[1.43/3] defines the "ghost field" ${\bf B}^{(3)}$
which is coupled by the relations [1.44] with the transversal  
components ${\bf B}^{(1)}$ and ${\bf B}^{(2)}$.\\ 

{\sc M.W. Evans}' {\bf B Cyclic Theorem} is the statement that each plane circularly 
polarized wave [1.43/1-2]
is accompanied by a longitudinal field [1.43/3], and the associated fields 
fulfil the cyclic equations [1.44].
{\sc M.W. Evans} considers this $O(3)$ hypothesis as a {\bf Law of Physics}.\\

\section{Checking the superposition property of the $O(3)$ hypothesis}

Instead of [1.38] we consider a phase shifted wave with the more general phase function

\begin{equation} 
\Phi_\alpha(t,z) = \omega t - \kappa z + \alpha = \Phi(t,z) + \alpha
\end{equation}

\noindent
which can be understood as a time shifted wave where the time shift is 
$t_0 := - \frac{\alpha}{\omega}$:

\begin{equation} 
\Phi_\alpha(t,z) = \Phi(t-t_0,z) .
\end{equation}

\noindent
We use the phase $\Phi_\alpha$  in [1.43] to obtain the time-shifted magnetic field

\begin{equation} 
{\bf B}^{(1)}
= \frac{1}{\sqrt{2}} B^{(0)}  
({\bf i}-i{\bf j}) e^{i(\Phi+\alpha)} ,
\end{equation}

\begin{equation} 
{\bf B}^{(2)}
= \frac{1}{\sqrt{2}} B^{(0)}   
({\bf i}+i {\bf j}) e^{-i(\Phi+\alpha)},
\end{equation}

\begin{equation} 
{\bf B}^{(3)} 
=
\gamma  B^{(0)} {\bf k} ,
\end{equation}

\noindent
where we have introduced a coefficient $\gamma$ that should equal $1$
following {\sc M.W. Evans} while in classical electrodynamics $\gamma =0$.\\

Now we consider the wave generated by the superposition of two waves with
the phase functions $\Phi_\alpha$  and $\Phi_{-\alpha}$,
respectively, and $\alpha$ such that $\cos{\alpha} < 1$. 
According to the {\em superposition principle} the total field is then 

\begin{equation} 
{\bf B}^{(1)} = \frac{1}{\sqrt{2}} B^{(0)}   
({\bf i}-i {\bf j}) [e^{i (\Phi+\alpha)} + e^{i (\Phi-\alpha)}]
= \frac{1}{\sqrt{2}} B^{(0)}   
({\bf i}-i {\bf j}) e^{i \Phi} 2 \cos{\alpha} ,
\end{equation}

\begin{equation} 
{\bf B}^{(2)} = \frac{1}{\sqrt{2}} B^{(0)}   
({\bf i}+i {\bf j}) [e^{-i(\Phi+\alpha)} + e^{-i(\Phi-\alpha)}]
= \frac{1}{\sqrt{2}} B^{(0)}   
({\bf i}+i {\bf j}) e^{-i \Phi} 2 \cos{\alpha} ,
\end{equation}

\begin{equation} 
{\bf B}^{(3)} = 2 \gamma  B^{(0)}  {\bf k} .
\end{equation}

Considering the first two (transversal) components we recognize that the 
superposition yields the original wave [1.43/1-2] multiplied by the factor 
$2 \cos{\alpha}$. Hence, according to {\sc M.W. Evans'} $O(3)$
hypothesis [1.43/3] it should be accompanied by a longitudinal component 
$2 \gamma  \cos{\alpha} \cdot B^{(0)} {\bf k}$ with $\gamma =1$.
The superposition principle, however, yields  
${\bf B}^{(3)} = 2 \gamma  B^{(0)} {\bf k} $ (Eq.8). 
Since we assumed $\cos{\alpha} < 1$, this is a contradiction.
Only the classical case $\gamma=0$
is compatible with the superposition principle, and 
{\sc M.W. Evans}' "ghost field" cannot exist.\\

{\large {\sc M.W. Evans}' cyclical O(3)-hypothesis {\em is 
incompatible with the superposition principle of waves.}}\\

{\em Remark without detailed proof:} A consequence of this 
incompatibility is that for $\gamma \not= 0$ it is impossible 
to construct (by Fourier synthesis) localized packets of 
circularly polarized waves. The transverse components 
${\bf B}^{(1)}$ and ${\bf B}^{(2)}$ have an oscillatory phase in space 
and time and can therefore be chosen so as to interfere destructively
outside the desired packet. However, according to eq. (1.43/3), the 
longitudinal ${\bf B}^{(3)}$ field of each Fourier component 
is a constant vector in space and time. By superposing such 
vectors we can only obtain another constant vector. The  
${\bf B}^{(3)}$ field of a wave packet must either disappear or fill the 
whole universe at all times. This amplifies an objection 
raised by A. Lakhtakia [6] already in 1995.

\end{document}